# On Augmented Kohn-Sham Potential for Energy as a Simple Sum of Orbital Energies

Mel Levy[1,2,3,*] and Federico Zahariev[4,&]

1. Department of Chemistry, Duke University, Durham, North Carolina 27708, USA
2. Department of Physics, North Carolina A&T State University, Greensboro, North Carolina 27411, USA
3. Department of Chemistry and Quantum Theory Group, Tulane University, New Orleans, Louisiana 70118, USA
4. Department of Chemistry and Ames Laboratory, Iowa State University, Ames, Iowa 50011, USA

It has recently been observed [Phys. Rev. Lett. 113, 113002 (2014)] that the ground-state energy may be obtained directly as a simple sum of augmented Kohn-Sham orbital energies, where it was ascertained that the corresponding one-body shifted Kohn-Sham effective potential has appealing features. With this in mind, eigenvalue and virial constraints are deduced for approximating this potential.

* mlevy@tulane.edu
& fzahari@iastate.edu

## I. Introduction

We have recently observed [1] that the exact ground-state energy of interest may be obtained directly as a simple sum of augmented Kohn-Sham orbital energies when a universal density-dependent constant is added to the familiar Kohn-Sham effective potential. The resultant effective potential, $\bar{w}([\rho];\boldsymbol{r})$, has enticing continuity properties. Thus the use of $\bar{w}([\rho];\boldsymbol{r})$ represents a viable "direct energy" approach.

The shifted Kohn-Sham potential, $\bar{w}([\rho];\boldsymbol{r})$, must be approximated. For this purpose, we deduce two constraints involving $\bar{w}([\rho];\boldsymbol{r})$, an eigenvalue constraint and a virial constraint. The "eigenvalue constraint" follows from eigenvalue equations (4) and (6), while the "virial constraint" follows from virial expressions (12) and (13).

## II. Eigenvalue Expressions

Assume that we are interested in the ground-state energy, $E_{GS}$, and electron density, $\rho_{GS}$, of

$$H = T + V_{ee} + \sum_{i=1}^{N} v(r_i), \qquad (1)$$

where $T$ is the kinetic energy operator, $V_{ee}$ is the electron repulsion operator, and $v(r)$ is the one-body attractive multiplicative potential. We have recently observed that the ground-state energy may be obtained as a simple sum of augmented Kohn-Sham orbital energies when a density-dependent constant is added to the usual Kohn-Sham effective potential to form $\bar{w}([\rho];r)$. That is, with $\bar{w}([\rho];r)$, $E_{GS}$ is simply

$$E_{GS} = \sum_{i=1}^{N} \bar{\varepsilon}_i, \qquad (2)$$

where

$$\left[ -\frac{1}{2}\nabla^2 + v(r) + \bar{w}([\rho_{GS}];r) \right] \varphi_i(r) = \bar{\varepsilon}_i \varphi_i(r), \qquad (3)$$

where $\varphi_i(r)$ are the Kohn-Sham orbitals and $\bar{\varepsilon}_i$ are the augmented Kohn-Sham orbital energies.

It is our purpose in this section to observe that expressions (2) and (3) can be combined into a single eigenvalue equation. Specifically,

$$\bar{H}_{KS} \Phi_{KS} = E_{KS} \Phi_{KS}, \qquad (4)$$

where $\Phi_{KS}$ is a single determinant that is formed from the $\varphi_i(r)$, and where

$$\bar{H}_{KS} = T + \sum_{i=1}^{N} v(r_i) + \sum_{i=1}^{N} \bar{w}([\rho_{GS}];r). \qquad (5)$$

Equation (5) compares nicely to its many-body counterpart, which is of course

$$H_{GS} \Psi_{GS} = E_{GS} \Psi_{GS}. \qquad (6)$$

We utilize eigenvalue equations (4) and (6) in the next section.

### III. Eigenvalue Constraint for $\bar{w}([\rho];r)$

For simplicity of presentation assume that $\Phi_{KS}$ and $\Psi_{GS}$ are real. Multiply Eq. (4) by $\Psi_{GS}$ and integrate, and multiply Eq. (6) by $\Phi_{KS}$ and integrate. Obtain

$$\langle \Psi_{GS} | \bar{H}_{KS} | \Phi_{KS} \rangle = E_{GS} \langle \Psi_{GS} | \Phi_{KS} \rangle \qquad (7)$$

and

$$\langle \Phi_{KS} | H | \Psi_{GS} \rangle = E_{GS} \langle \Phi_{KS} | \Psi_{GS} \rangle. \qquad (8)$$

Subtract Eq. (7) from Eq. (8) to obtain

$$\langle \Phi_{KS} | \sum_{i=1}^{N} \bar{w}([\rho_{GS}];r) | \Psi_{GS} \rangle = \langle \Phi_{KS} | V_{ee} | \Psi_{GS} \rangle \qquad (9a)$$

or,

$$\frac{1}{N-1} \iint \{ \bar{w}([\rho_{GS}];r) + \bar{w}([\rho_{GS}];r') \} \rho_{KS,GS}(r,r') dr dr'$$
$$= \iint \frac{1}{|r-r'|} \rho_{KS,GS}(r,r') dr dr' \qquad, (9b)$$

where N is the number of electrons and $\rho_{KS,GS}(r;r')$ is the transition pair density that is obtained from the product of $\Phi_{KS}$ and $\Psi_{GS}$. Eq. (9) is our first desired constraint for $\bar{w}([\rho];r)$. Note that Eq. (9) involves nicely only $\bar{w}([\rho];r)$ and $V_{ee}$ as operators. We call Eq. (9) an "eigenvalue constraint" for $\bar{w}([\rho];r)$ because the constraint follows directly from eigenvalue equations (4) and (6).

### IV. Virial Constraint for $\bar{w}([\rho];r)$

$E_{GS}$ and $\rho_{GS}$ are connected by

$$E_{GS} = T_s[\rho_{GS}] + \int \left[ v(\mathbf{r}) + \bar{w}([\rho_{GS}];\mathbf{r}) \right] \rho_{GS}(\mathbf{r}) d\mathbf{r}, \quad (10)$$

where $T_s[\rho_{GS}]$ is the non-interacting Kohn-Sham kinetic energy functional. From the fact that the right-hand side of Eq. (10) is raised when $\rho_{GS}$ is scaled uniformly by a coordinate scale factor, at fixed $v(\mathbf{r}) + \bar{w}([\rho_{GS}];\mathbf{r})$, and from the fact that $T_s[\rho_{GS}]$ scales as the square of the scale factor [2] when $\rho_{GS}$ is scaled, it follows that

$$2T_s[\rho_{GS}] = -\int \left[ v(\mathbf{r}) + \bar{w}([\rho_{GS}];\mathbf{r}) \right] \left[ 3\rho_{GS}(\mathbf{r}) + \mathbf{r} \cdot \nabla \rho_{GS}(\mathbf{r}) \right] d\mathbf{r}. \quad (11)$$

The combination of expressions (10) and (11) gives the following virial expression involving $\bar{w}([\rho];\mathbf{r})$:

$$E_{GS} = -\frac{1}{2} \int \left[ v(\mathbf{r}) + \bar{w}[\rho_{GS}](\mathbf{r}) \right] \left[ \rho_{GS}(\mathbf{r}) + \mathbf{r} \cdot \nabla \rho_{GS}(\mathbf{r}) \right] d\mathbf{r}. \quad (12)$$

Next, compare Eq. (12) with the familiar virial expression for $E_{GS}$ that involves $V_{ee}$ and arises from the fact that $\langle H \rangle$ is raised when the coordinates of $\Psi_{GS}$ are scaled. This virial expression is

$$E_{GS} = \frac{1}{2} \langle \Psi_{GS} | V_{ee} | \Psi_{GS} \rangle - \frac{1}{2} \int v(\mathbf{r}) \left[ \rho_{GS}(\mathbf{r}) + \mathbf{r} \cdot \nabla \rho_{GS}(\mathbf{r}) \right] d\mathbf{r}. \quad (13)$$

The comparison of equation (12) and (13) gives

$$\langle \Psi_{GS} | V_{ee} | \Psi_{GS} \rangle = -\int \bar{w}([\rho_{GS}];\mathbf{r}) \left[ \rho_{GS}(\mathbf{r}) + \mathbf{r} \cdot \nabla \rho_{GS}(\mathbf{r}) \right] d\mathbf{r} \quad (14a)$$

or,

$$\frac{1}{2} \iint \frac{\rho_{GS}(\mathbf{r},\mathbf{r}')}{|\mathbf{r}-\mathbf{r}'|} d\mathbf{r} d\mathbf{r}' = -\int \bar{w}([\rho_{GS}];\mathbf{r}) \left[ \rho_{GS}(\mathbf{r}) + \mathbf{r} \cdot \nabla \rho_{GS}(\mathbf{r}) \right] d\mathbf{r}, \quad (14b)$$

where $\rho_{GS}(\mathbf{r},\mathbf{r}')$ is the ground-state pair density. Equation (14) is our second desired constraint for $\bar{w}([\rho];\mathbf{r})$.

V. Discussion

We have deduced the "eigenvalue constraint" for $\bar{w}([\rho];\boldsymbol{r})$, which is expression (9), and the "virial constraint" for $\bar{w}([\rho];\boldsymbol{r})$, which is expression (14). In practice, to test an approximate $\bar{w}([\rho];\boldsymbol{r})$, both constraints are intended for use with interacting wavefunctions from accurate calculations. For understanding $\bar{w}([\rho];\boldsymbol{r})$, it is helpful that $V_{ee}$ is the only operator involved in both constraints, because the non-interacting $\bar{w}([\rho];\boldsymbol{r})$ is meant to incorporate indirectly much of the interacting nature of $V_{ee}$.

We close by noting that $\lim_{|\boldsymbol{r}|\to\infty}\bar{w}([\rho];\boldsymbol{r})$ is important because it contains significant energy information. With this in mind, $\lim_{|\boldsymbol{r}|\to\infty}\bar{w}([\rho];\boldsymbol{r})$ has recently been determined exactly in the strongly correlated regime [3], and an exchange potential has been recently put forth [4] that incorporates naturally an approximation for $\lim_{|\boldsymbol{r}|\to\infty}\bar{w}([\rho];\boldsymbol{r})$. Also, a significant practical test for the approximation of $\lim_{|\boldsymbol{r}|\to\infty}\bar{w}([\rho];\boldsymbol{r})$ is the requirement that the value of the total energy from an output density must not be higher than the value from the corresponding input density at each iteration towards self-consistency [5]. This constraint embodies several exact properties of the potential. Further studies include electron number changing density scaling [6] and different ways of adding constants to the potential [7].

It is a pleasure to dedicate this paper to Andreas Savin.